%% file: main.tex
\documentclass[10pt,conference]{IEEEtran}
\IEEEoverridecommandlockouts
\input{package}

\input{macro}

\begin{document}

\title{A Hierarchical and Evolvable Benchmark for Fine-Grained Code Instruction Following with Multi-Turn Feedback}

\author{
    \IEEEauthorblockN{
        Guoliang Duan\IEEEauthorrefmark{1},
        Mingwei Liu\IEEEauthorrefmark{1}\textsuperscript{*},
        Yanlin Wang\IEEEauthorrefmark{1},
        Chong Wang\IEEEauthorrefmark{2},\\
        Xin Peng\IEEEauthorrefmark{3},
        Zibin Zheng\IEEEauthorrefmark{1}
    }
    \IEEEauthorblockA{\IEEEauthorrefmark{1}Sun Yat-sen University, Zhuhai, China\\
    duangliang@mail2.sysu.edu.cn, \{liumw26, wangylin36, zhzibin\}@mail.sysu.edu.cn}
    \IEEEauthorblockA{\IEEEauthorrefmark{2}Nanyang Technological University, Singapore\\
    chong.wang@ntu.edu.sg}
    \IEEEauthorblockA{\IEEEauthorrefmark{3}Fudan University, Shanghai, China\\
    pengxin@fudan.edu.cn}
    \thanks{\textsuperscript{*}Corresponding author.
    }
}
\maketitle

\begin{abstract}
    \input{abstract}
\end{abstract}


\section{Introduction}
\label{sec:intro}
\input{sections/introduction}


\input{sections/taxonomy}

\section{Benchmark Construction} 
\label{sec:data_construction}
\input{sections/data_construction}

\section{Experiments}
\label{sec:experiments}
\input{sections/experiments}

\section{Related Work}
\input{sections/related_work}

\section{Threats}
\label{sec:threats}
\input{sections/threats}

\section{Conclusions}
\input{sections/conclusion}

\normalem
\bibliographystyle{IEEEtran}
\balance
\bibliography{refs}

\end{document}

%% file: package.tex
\usepackage{booktabs}
\usepackage{multirow}
\usepackage{algorithmic}
\usepackage{graphicx}
\usepackage{textcomp}
\usepackage{xcolor}
\usepackage[hidelinks]{hyperref}

\usepackage{makecell}
\usepackage{tabularx}
\usepackage[T1]{fontenc}
\usepackage{comment}
\usepackage{xcolor}
\usepackage{multirow}
\usepackage{array}
\usepackage{soul}
\usepackage{threeparttable}
\usepackage{balance}
\usepackage{diagbox}
\usepackage{ulem}
\usepackage{subfigure}
\usepackage{microtype} 
\usepackage{calc}

\usepackage{enumitem}
\usepackage{url}
\usepackage{xspace}
\usepackage{float}
\usepackage{stfloats}
\usepackage{adjustbox}
\usepackage{amsmath,amssymb,amsfonts}
\usepackage{algorithmic}
\usepackage{graphicx}
\usepackage{textcomp}
\usepackage{xcolor}
\usepackage{multirow}
\usepackage{mdframed}
\usepackage{tcolorbox}
\usepackage{makecell}
\usepackage{threeparttable}
\usepackage{colortbl}
\usepackage{subfigure}
\usepackage{hhline}
\usepackage{pifont}
\usepackage{enumitem}

\usepackage{amssymb}

%% file: macro.tex
\newcommand\app{ConstraGen\xspace}
\newcommand\bench{MultiCodeIF\xspace}

\newcommand\todo[1]{{\color{blue} \textbf{TODO: }\color{blue}\color{blue}#1}}

\newcommand{\parabf}[1]{\noindent\textbf{#1}}

\newcommand{\Comment}[1]{}

\definecolor{ggray}{HTML}{eff0f0}
\definecolor{gggray}{HTML}{E8E8E8}
\definecolor{ggggray}{HTML}{BEBEBE}

\usepackage[utf8]{inputenc}
\usepackage[english]{babel}

\usepackage{enumitem} 

\newcounter{finding}

\newcommand{\distance}{5pt}

%% file: abstract.tex
Large language models (LLMs) have advanced significantly in code generation, yet their ability to follow complex programming instructions with layered and diverse constraints remains underexplored. Existing benchmarks often prioritize functional correctness, overlooking the nuanced requirements found in real-world development. 

We introduce \bench{}, a comprehensive benchmark designed to evaluate instruction-following in code generation across multiple dimensions: constraint type, hierarchical levels, and iterative refinement. Built upon a structured taxonomy of 9 categories and 27 constraint types, \bench{} enables granular assessment of both functional and non-functional instruction adherence. Using an automated pipeline, \app{}, we synthesize and evolve 2,021 code tasks sourced from 14 programming languages, supporting multi-turn evaluation through feedback-driven task variants.

Empirical evaluation of six state-of-the-art LLMs uncovers substantial performance disparities. The top-performing model, Claude-3-7-Sonnet, achieves 63.0\% average constraint satisfaction, while smaller models like Qwen3-1.7B fall to 44.8\%. Models perform well on explicit constraints (over 70\%), but struggle with implicit or abstract constraints (below 40\%). Tasks with multiple hierarchical constraints significantly reduce model success rates, from 54.5\% in single-level to just 18.8\% in multi-level scenarios. However, structured feedback enables progressive improvement: average constraint satisfaction rises from 63.0\% to 83.4\% over four iterative refinement rounds.
\bench{} provides a scalable, constraint-aware, and feedback-sensitive framework to benchmark LLMs under realistic code generation scenarios, bridging the gap between synthetic evaluations and real-world instruction complexity. The full benchmark dataset, evaluation pipeline, and source code are available at \url{https://github.com/SYSUSELab/MultiCodeIF}.

%% file: sections/introduction.tex
With the rapid advancement of large language models (LLMs) in code generation, their capabilities in automating programming tasks have attracted increasing attention~\cite{lin2024soen101codegenerationemulating,jiang2025rocodeintegratingbacktrackingmechanism,10.1109/ICSE48619.2023.00179,di2025enhancingcodegenerationbidirectional,tian2024fixinglargelanguagemodels}. General-purpose models such as GPT-4 have demonstrated strong potential in understanding and executing programming instructions~\cite{openai2024gpt4technicalreport}, accelerating the deployment of intelligent code generation tools across both academia and industry.

However, existing evaluation frameworks for code generation primarily focus on functional correctness or code completion accuracy, for example Pass@1 at HumanEval \cite{chen2021evaluatinglargelanguagemodels} and MBPP \cite{austin2021programsynthesislargelanguage}, two popular code generation benchmark. 
They rarely assess whether models can faithfully understand and follow natural language instructions that contain nuanced, fine-grained constraints~\cite{chen2025surveyevaluatinglargelanguage,wang2025codeifbenchevaluatinginstructionfollowingcapabilities,yan2025codeifbenchmarkinginstructionfollowingcapabilities}. 
This results in a gap between current evaluation practices and the real-world needs of software development.

\begin{figure}
    \centering
    \includegraphics[width=\linewidth]{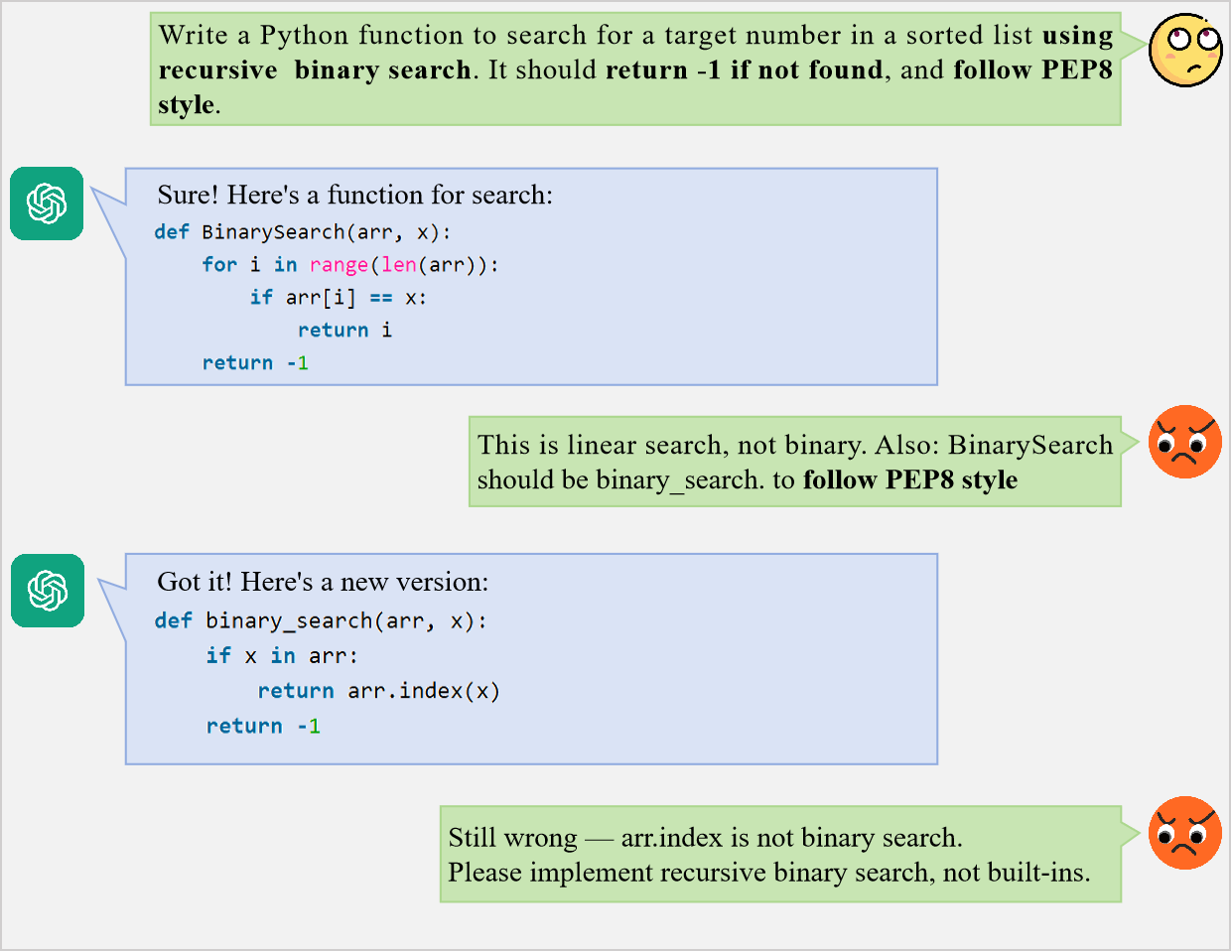}
    \caption{LLM's Failure in Multi-Constraint Code Generation}
    \label{fig:motivation_scenario}
\end{figure}

Figure~\ref{fig:motivation_scenario} shows an LLM repeatedly failing to complete a code generation task with multi-level constraints: it first violates both the algorithm and naming requirements, then partially fixes the naming but still gets the logic wrong. This reflects a common issue, LLMs often struggle to satisfy multiple, diverse constraints simultaneously, tending to address only part of the instruction at each step. Even with explicit feedback, the model fails to fully recover, indicating that iterative prompting alone cannot ensure full compliance. This highlights the need for fine-grained, constraint-aware, and feedback-sensitive benchmarks that go beyond functional correctness to assess instruction-following under realistic conditions.

Although several benchmarks have been proposed for evaluating LLMs’ instruction-following abilities in general NLP tasks, such as summarization, question answering, and dialogue, the code generation domain remains less underexplored in this regard. 
Table~\ref{table:comp_existing_datasets} compares representative instruction-following benchmarks. 
While benchmarks like IFEval~\cite{IFEval} and InFoBench~\cite{InFoBench} capture natural language instructions, they are typically task-agnostic and lack code-specific constraint modeling. 
More recent efforts such as CodeIF~\cite{CodeIF} and CodeIF-bench~\cite{CodeIF-Bench} attempt to include coding constraints but still suffer from limitations: coarse constraint granularity, limited language support, static task definitions, and no modeling of iterative correction or self-repair processes.

\input{tables/datasets_comp}

To address these gaps, we introduce \bench, a new benchmark tailored for evaluating LLM instruction-following capabilities in code generation scenarios. 
\bench{} is designed to be: \textbf{(1) code-specific}, explicitly grounded in programming tasks and semantics; \textbf{(2) constraint-centric}, supporting a diverse and fine-grained constraint taxonomy across functional and non-functional dimensions; and \textbf{(3) evolvable and interactive}, featuring an LLM-driven task generation pipeline that enables continuous expansion and reduces the risk of data leakage, while also supporting multi-round refinement scenarios to assess how models respond to feedback and progressively satisfy complex constraints.

\bench is built upon a comprehensive constraint taxonomy that includes 9 high-level categories and 27 fine-grained types. This taxonomy spans a wide spectrum of programming-relevant constraints, from language syntax and API usage to algorithmic complexity, parameter types, naming conventions, contextual cues, and real-world development scenarios. It unifies hard constraints that affect correctness and soft constraints that shape usability, readability, or realism, enabling fine-grained, multifaceted evaluation of LLM outputs. To ensure coverage, rigor, and operationalizability, each constraint type in the taxonomy is precisely defined, supported by rule-based or LLM-based evaluation strategies.

To efficiently construct \bench, we propose an automatic pipeline \app . Starting from real-world seed code and extracted programming concepts, we define a structured constraint system based on a hierarchical taxonomy and leverage advanced LLMs to synthesize constraint-aligned tasks via prompt engineering. To ensure diversity and correctness, we apply similarity-based filtering and manual validation. Tasks are further evolved into multi-constraint variants through LLM-guided augmentation, enabling controlled complexity growth. Additionally, \app{} includes iterative correction scenarios to evaluate whether LLMs can recover from constraint violations after one or multiple feedback rounds.

The resulting \bench{} includes 1,526 single-constraint and 495 multi-constraint tasks, each associated with explicit, fine-grained constraints. It spans diverse programming scenarios across 14 languages (e.g., Python, Java, C++, Rust) and supports scalable task refinement for longitudinal evaluation of LLMs over multiple interaction rounds.

Using \bench, we evaluate six state-of-the-art LLMs and analyze their instruction-following behaviors under diverse and evolving constraints. Our experiments reveal key findings:
(1) Instruction adherence varies widely; top model \textbf{Claude-3-7-Sonnet} achieves 63.0\% average constraint satisfaction, while smaller open-source Qwen3-1.7B scores 44.8\%, a ~20 point gap. 
(2) Constraint difficulty differs: concrete types like \textbf{Environment} and \textbf{Code Context} exceed 70\% satisfaction, while abstract ones like \textbf{Code Quality} fall below 40\%.
(3) Multi-level constraints pose challenges; hard satisfaction rate (HSR) drops from 54.5\% (single) to 18.8\% (multi-constraint), with errors compounding as complexity grows.
(4) Iterative self-repair through structured feedback boosts satisfaction from 63.0\% to 83.4\% over four rounds, highlighting the promise of feedback-driven refinement.

The contributions of this paper are as follows:
\begin{itemize}
\item A hierarchical constraint taxonomy formalizing multi-level instruction adherence in code generation.
\item Construction of \bench, a large-scale, multi-type, multi-level, multi-turn code instruction following benchmark, publicly available~\cite{replication_package}.
\item A scalable pipeline, \app,  for generating and evolving instruction tasks using open-source code and LLMs.
\item Comprehensive evaluation of six LLMs on \bench, offering insights into their strengths and limitations in complex instruction following.
\end{itemize}

%% file: tables/datasets_comp.tex
\begin{table*}[]
\scriptsize
\renewcommand{\arraystretch}{0.9}
\label{table:comp_existing_datasets}
\centering
\caption{Comparison of Existing Instruction-Following Benchmarks}
\begin{tabular}{@{}l|ccccccc@{}}
\toprule
\textbf{Dataset} &
  \textbf{Tasks} &
  \textbf{Code Domain} &
  \textbf{\begin{tabular}[c]{@{}c@{}}Fine-Grained\\ Constraints\end{tabular}} &
  \textbf{Supported Languages} &
  \textbf{Multi-level} &
  \textbf{Evolvable} &
  \textbf{Self-repair} \\ \midrule
IFEval~\cite{IFEval}       & 541  & $\times$     & $\times$     & -                     & $\times$     & $\times$     & $\times$ \\
InFoBench~\cite{InFoBench}    & 500  & $\times$     & $\checkmark$ & -                     & $\times$     & $\times$     & $\times$ \\
FollowBench~\cite{FollowBench}  & 820  & $\times$     & $\checkmark$ & -                     & $\checkmark$ & $\times$     & $\times$ \\
CodeIF-Bench~\cite{CodeIF-Bench} & 879  & $\checkmark$ & $\times$     & Python                & $\times$     & $\times$     & $\times$ \\
CodeIF~\cite{CodeIF}       & 1,200 & $\checkmark$ & $\checkmark$ & Python, Java, Go, C++ & $\times$     & $\times$     & $\times$ \\ \midrule
\bench &
  2,021 &
  $\checkmark$ &
  $\checkmark$ &
  \begin{tabular}[c]{@{}c@{}}Python, JavaScript, etc. (14 in total)\end{tabular} &
  $\checkmark$ &
  $\checkmark$ &
  $\checkmark$ \\ \bottomrule
\end{tabular}
\end{table*}

%% file: sections/taxonomy.tex
\section{Fine-Grained Constraint Taxonomy}
\label{sec:taxonomy}

To support rigorous evaluation of instruction-following in code generation, we introduce a fine-grained taxonomy of constraints tailored to programming tasks (Table~\ref{tab:constraint-types}). It includes both explicit constraints, such as syntax, data structures, and algorithmic requirements, and implicit ones that reflect common yet unstated expectations, such as alignment with code context or realistic development scenarios. These constraints vary in impact, influencing both functional correctness and non-functional qualities like readability or robustness.

We construct the taxonomy in three steps. First, we build on prior work, such as the 24 coding style violations identified by Wang et al. \cite{wang2024beyond} and the five broad constraint types proposed by FollowBench \cite{jiang2023followbench}. Second, we qualitatively analyze real-world prompts from ShareGPT~\cite{ShareGPT} to identify naturally occurring constraint patterns. Third, we synthesize these insights into a taxonomy with 9 top-level categories and 27 fine-grained types, each linked to practical evaluation strategies, ranging from static rule-based checks (e.g., API usage) to LLM-based assessments (e.g., judging code clarity via prompt-based evaluation~\cite{ballesteroribó2025promptbasedcosteffectiveevaluationoperation}).

\input{tables/constraint_types}

Table~\ref{tab:constraint-types} summarizes the full taxonomy. Below are brief descriptions of the nine top-level categories:
\begin{itemize}
    \item \textbf{Interface Specification:} Constraints on inputs and outputs, including parameter types, counts, and returns.
    \item \textbf{Environment:} Limits on languages, versions, frameworks, libraries, APIs, or language features.
    \item \textbf{Data Structure:} Requirements for data structure types, sizes, or operations.
    \item \textbf{Algorithm:} Specifications of algorithm types or performance (time/space complexity).
    \item \textbf{Coding Style:} Rules for naming, indentation, braces, comments, and formatting.
    \item \textbf{Code Quality:} Expectations for readability, maintainability, and robustness.
    \item \textbf{Scenario:} Contextual constraints reflecting application environment or usage.
    \item \textbf{Code Context:} Ensures code aligns with or fits existing surrounding code.
    \item \textbf{Exemplar:} Follow or mimic example code’s structure, logic, or style.
\end{itemize}

Prior studies offer useful foundations but often lack code-specific detail, comprehensive scope, or practical evaluation methods. For example, FollowBench~\cite{jiang2023followbench} proposes a general five-category system not tailored to code; Wang et al.~\cite{wang2024beyond} focus on style violations but omit semantic and contextual constraints; CodeIF~\cite{yan2025codeifbenchmarkinginstructionfollowingcapabilities} addresses control structures without clear evaluation strategies. Our taxonomy fills these gaps by providing (1) precise constraint definitions, (2) broad coverage of both functional and non-functional properties, and (3) concrete evaluation methods ranging from rule-based checks to LLM-based assessments. It is extensible to incorporate new constraint types and adapt to evolving code generation needs.

In summary, this taxonomy offers a practical, scalable framework for instruction design, evaluation, and behavior control in real-world code generation.

%% file: tables/constraint_types.tex
\begin{table*}[]
\scriptsize
\centering
\caption{Overview of constraint types in our benchmark with examples}
\label{tab:constraint-types}
\small
\begin{tabular}{|c|l|c|l|c|l|}
\hline
\textbf{ID} &
  \multicolumn{1}{c|}{\textbf{Typing}} &
  \textbf{Category} &
  \textbf{Constraint Type} &
  \textbf{\begin{tabular}[c]{@{}c@{}}Evaluation\end{tabular}} &
  \textbf{Example} \\ \hline
1 &
  \multirow{24}{*}{Explicit} &
  \multirow{6}{*}{\begin{tabular}[c]{@{}c@{}}Interface\\ Specification\end{tabular}} &
  Parameter Type &
  Rule &
  using a dictionary as parameter \\ \cline{1-1} \cline{4-6} 
2 &
   &
   &
  Parameter Range &
  LLM &
  the integer parameter must be non-negative \\ \cline{1-1} \cline{4-6} 
3 &
   &
   &
  Parameter Signature &
  Rule &
  parameter signature must be {[}String, boolean{]} \\ \cline{1-1} \cline{4-6} 
4 &
   &
   &
  Return Type &
  Rule &
  function must return a list \\ \cline{1-1} \cline{4-6} 
5 &
   &
   &
  Return Range &
  LLM &
  the returned integer must be non-negative \\ \cline{1-1} \cline{4-6} 
6 &
   &
   &
  Return Cardinality &
  Rule &
  return cardinality must be a tuple (map, error) \\ \cline{1-1} \cline{3-6} 
7 &
   &
  \multirow{6}{*}{Environment} &
  Language Type &
  Rule &
  using Python. \\ \cline{1-1} \cline{4-6} 
8 &
   &
   &
  Language Version &
  Rule &
  using Java version JDK11 \\ \cline{1-1} \cline{4-6} 
9 &
   &
   &
  Advanced Syntax &
  Rule &
  using lambda expression in Kotlin \\ \cline{1-1} \cline{4-6} 
10 &
   &
   &
  Function/Method Invocation &
  Rule &
  using string.len() function \\ \cline{1-1} \cline{4-6} 
11 &
   &
   &
  API/Library Usage &
  Rule &
  using pandas library \\ \cline{1-1} \cline{4-6} 
12 &
   &
   &
  Framework &
  LLM &
  using Flask Web framework \\ \cline{1-1} \cline{3-6} 
13 &
   &
  \multirow{3}{*}{Data Structure} &
  Data Structure Type &
  Rule &
  using a BinarySearchTree data structure \\ \cline{1-1} \cline{4-6} 
14 &
   &
   &
  Data Structure Scale &
  Rule &
  the list cannot contain more than 100 elements \\ \cline{1-1} \cline{4-6} 
15 &
   &
   &
  Data Structure Operation &
  Rule &
  must use the pop() and top() operation in stack \\ \cline{1-1} \cline{3-6} 
16 &
   &
  \multirow{3}{*}{Algorithm} &
  Algorithm Type &
  LLM &
  using dp / divide and conquer algorithm \\ \cline{1-1} \cline{4-6} 
17 &
   &
   &
  Algorithm Time Complexity &
  LLM &
  ensure O(1) time complexity \\ \cline{1-1} \cline{4-6} 
18 &
   &
   &
  Algorithm Space Complexity &
  LLM &
  ensure O(1) space complexity \\ \cline{1-1} \cline{3-6} 
19 &
   &
  \multirow{5}{*}{Coding Style} &
  Naming Convention &
  Rule &
  using snake\_case for variables \\ \cline{1-1} \cline{4-6} 
20 &
   &
   &
  Indentation Style &
  Rule &
  using `\textbackslash{}t' for indentation \\ \cline{1-1} \cline{4-6} 
21 &
   &
   &
  Brace Style &
  Rule &
  using K\&R / Allman brace style \\ \cline{1-1} \cline{4-6} 
22 &
   &
   &
  Comment Style &
  Rule &
  using docstring for each function \\ \cline{1-1} \cline{4-6} 
23 &
   &
   &
  Declaration Style &
  LLM &
  variables be declared at the top of a function \\ \cline{1-1} \cline{3-6} 
24 &
   &
  Code Quality &
  Code Quality Type &
  LLM &
  ensuring code readability requirements \\ \hline
25 &
  \multicolumn{1}{c|}{\multirow{3}{*}{Implicit}} &
  Scenario &
  Scenario Type &
  LLM &
  in the algorithm competition scene... \\ \cline{1-1} \cline{3-6} 
26 &
  \multicolumn{1}{c|}{} &
  Code Context &
  Concrete Code Context &
  LLM &
  \{concrete\_code\} ... using the method in the code \\ \cline{1-1} \cline{3-6} 
27 &
  \multicolumn{1}{c|}{} &
  Exemplar &
  Concrete Code Exemplar &
  LLM &
  \{concrete\_code\} ... following the example code \\ \hline
\end{tabular}
\end{table*}

%% file: sections/data_construction.tex
\begin{figure*}[htb]
	\centering
	\includegraphics[width=1.5\columnwidth]{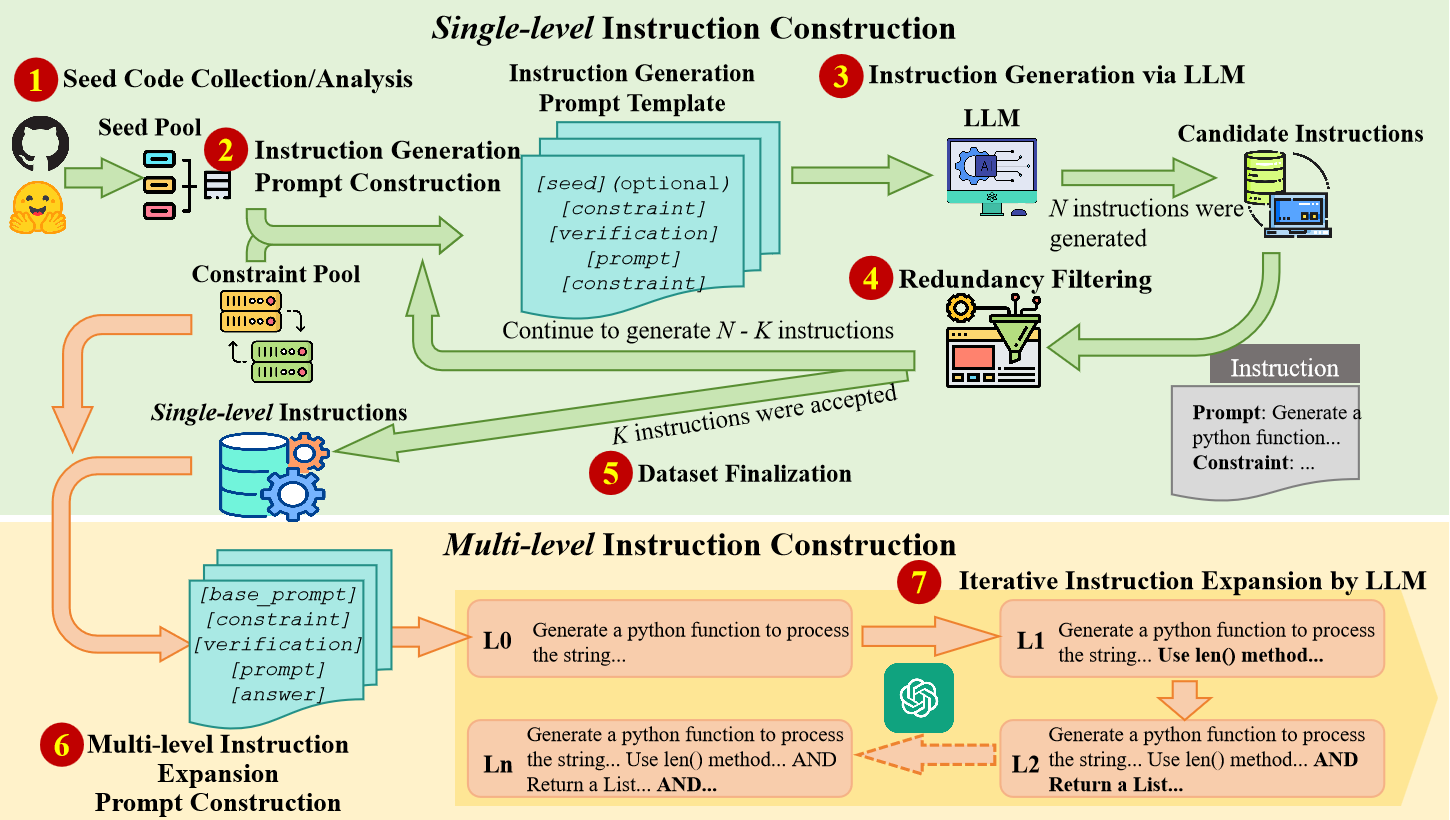}
	\caption{Overview of \app}
    
	\label{fig:data_construction_overview}
\end{figure*}

To construct \bench, we propose \app, a fully automated pipeline for generating diverse, fine-grained instruction-following tasks for code generation models. Based on a taxonomy of constraints and open-source code, \app comprises two stages (Figure~\ref{fig:data_construction_overview}): \textbf{(1) single-level instruction construction}, where LLMs generate tasks conditioned on one constraint; and \textbf{(2) multi-level instruction expansion}, where additional constraints are incrementally composed to raise task difficulty. The pipeline integrates prompt templating, constraint sampling, LLM generation, redundancy filtering, and iterative refinement, enabling scalable and systematic evaluation of LLMs under increasing complexity.

We introduce the dataset format, followed by the single-level instruction construction, multi-level instruction expansion, and benchmark statistics.

\subsection{Data Format}
\label{sec:data_construction:data_format}
\input{sections/data_format}

\input{sections/single-level-construction}

\input{sections/muti-level-expansion}

\subsection{Benchmark Statistics}
\label{sec:app:dataset_overview}
\input{sections/dataset_overview}


%% file: sections/data_format.tex
\begin{figure}[tbh]
	\centering
	\includegraphics[width=0.65\columnwidth]{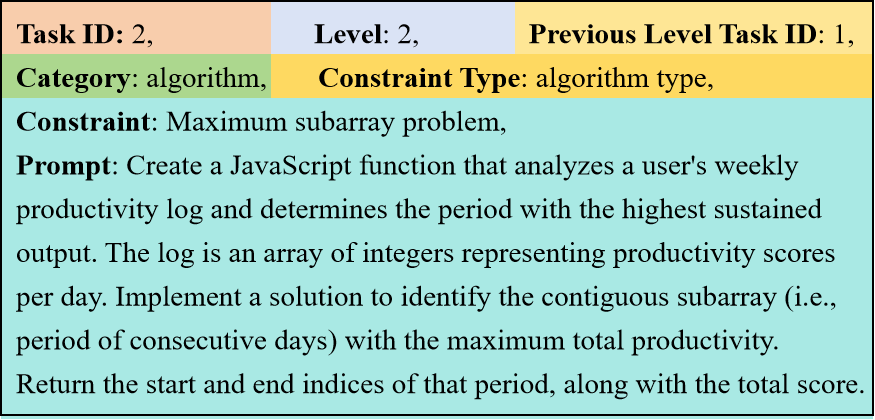}
	\caption{An Example of an L2 (Level 2) Task in \bench}
	\label{fig:data_format_example}
\end{figure}

The code generation task in our benchmark is represented as a structured instance: \textit{<Task ID, Level, Previous Level Task ID, Category, Constraint Type, Constraint, Prompt>}. As shown in Figure~\ref{fig:data_format_example}, the \textit{Category} and \textit{Constraint Type} are derived from the constraint taxonomy in Table~\ref{tab:constraint-types}. The \textit{Task ID} uniquely identifies each task, while the \textit{Previous Level Task ID} links it to its parent in the constraint hierarchy. The \textit{Constraint} specifies the concrete requirement to be satisfied, and the \textit{Prompt} provides the natural language instruction for code generation. The example shown in Figure~\ref{fig:data_format_example} is an L2-level task, which is based on an L1-level task (i.e., only one constraint) and adds an additional algorithm constraint. This structured format enables fine-grained constraint control and supports precise evaluation of model outputs in terms of correctness and constraint adherence.

%% file: sections/single-level-construction.tex
\subsection{Single-level Instruction Construction}
\label{sec:data_construction:single-level}
The stage focuses on generating instruction-following tasks with exactly one constraint (Level 1, or $L1$). This controlled complexity ensures clarity in evaluating how well models comply with individual constraints. The pipeline comprises five key steps, detailed below:

\parabf{Step 1: Seed Code Collection/Analysis .} We start by collecting a diverse set of seed code snippets from various open-source repositories, such as GitHub and Hugging Face. These snippets span multiple programming languages and domains to capture real-world coding diversity and ensure broad task coverage.
To semantically abstract these snippets, we utilize a powerful LLM (GPT-4-Turbo ~\cite{austin2021programsynthesislargelanguage} in our implementations) to analysis them to extract core programming concepts, such as algorithmic patterns or design principles, embedded within the code. Pairing each snippet with its extracted concepts (i.e., <code\_snippet, concepts>) serves two purposes: it informs subsequent prompt construction with meaningful context and enables the generation of semantically rich, constraint-aligned tasks. This approach strengthens the relevance and diversity of generated instructions, crucial for robust benchmarking.

\begin{figure}[htbp]
    \centering
    \includegraphics[width=0.9\linewidth]{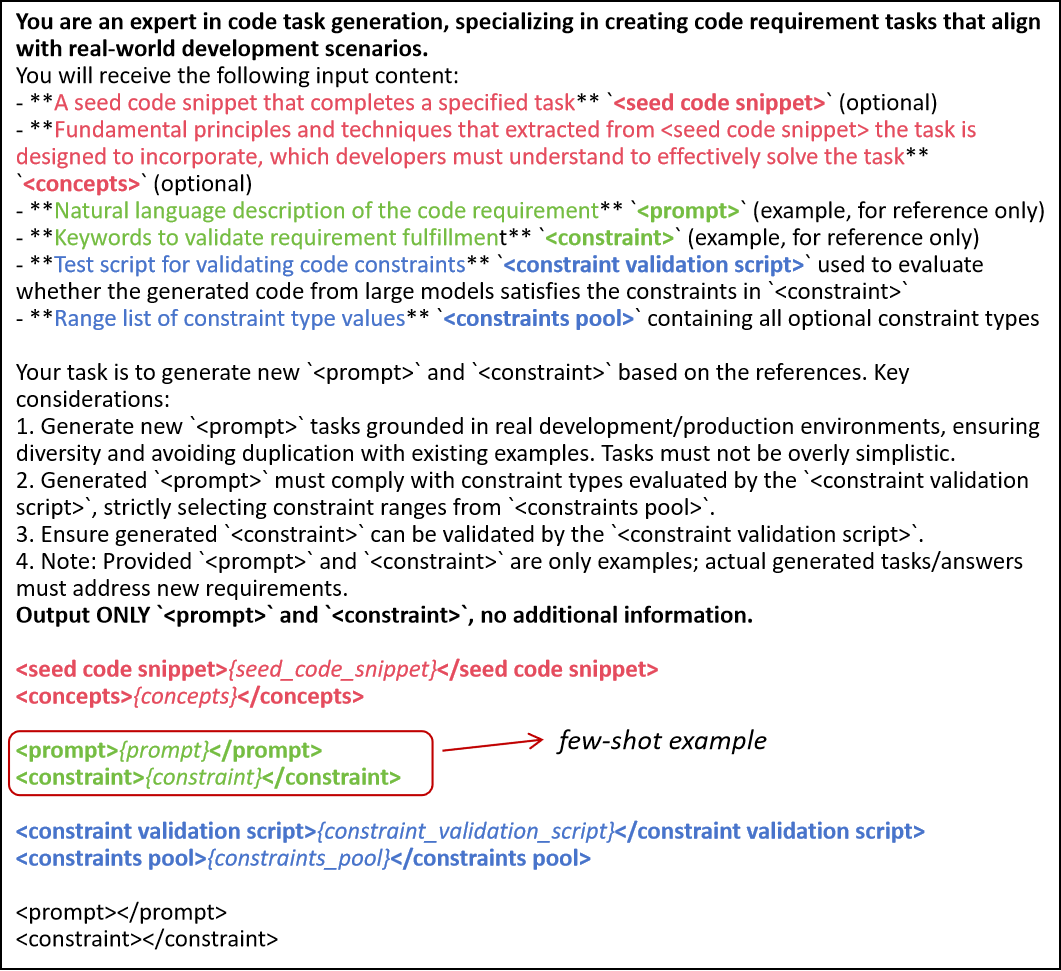} 
    \caption{Structured prompt template for code instruction generation. It consists of six components: \textbf{seed code snippet} (a collected sample code snippet used to inform task design), \textbf{concepts} (core programming principles extracted from the seed code), \textbf{prompt} (a natural language description of code generation task), \textbf{constraint} (keywords representing the constraints the generated code must satisfy), \textbf{constraint validation script} (a script for automatically evaluating constraint satisfaction), and \textbf{constraint pool} (a predefined set of constraints).}
\label{fig:prompt_template}
\end{figure}

\parabf{Step 2: Instruction Generation Prompt Construction.}
For each constraint type, we assemble a generation prompt tailored to that specific category. Figure~\ref{fig:prompt_template} show the prompt template. The prompt integrates a seed code snippet, its associated concepts, relevant few-shot examples, and a constraint sampled from the corresponding constraint pool, and constraint Validation scripts. Notably, each fine-grained constraint type (e.g., language type, parameter type, algorithm type) is associated with its own specialized constraint pool, for instance, the language type constraint pool includes values such as Java, Python, and Rust.

To systematically guide task generation, we establish a structured prompt engineering framework composed of three key components (see Figure~\ref{fig:prompt_template}):
\begin{itemize}
    \item Constraint Validation Scripts: Automated scripts that verify whether the generated code satisfies specified constraints. These scripts form the backbone of the evaluation pipeline, enabling objective, reproducible, and automated assessment of constraint compliance. For example, Data Structure validation script uses static analysis tools like Tree-sitter~\cite{max_brunsfeld_2025_15531772} to parse code to confirm compliance with the required constraint. Code Quality validation uses a reference LLM to judge whether the code's logic aligns with the specified instruction.
    \item Constraint Pool: A curated taxonomy of constraint categories, accompanied by detailed descriptions and parameter ranges. This structured pool allows precise and scalable constraint specification, while the contextual scenarios embedded within help the LLM generate relevant and well-aligned instructions. For example, the Environment category includes constraints such as "\textit{must use JDK11 for Java version}" or "\textit{implement using a lambda expression}", each annotated with formal definitions and typical usage scenarios.
    \item Prompt-Constraint Pairs: A curated set of exemplar tasks, including natural language prompts and verified constraint-compliant code solutions. These pairs serve as few-shot examples during LLM-based generation, enhancing output quality and alignment via in-context learning.
\end{itemize}

This modular design strikes a balance between flexibility and control, facilitating the creation of diverse, constraint-driven code generation tasks while ensuring consistency and high quality. Note all constraint validation scripts, constraints in constraint pool, and  Prompt-Constraint Pairs are included in our replication package~\cite{replication_package}.

\parabf{Step 3: Instruction Generation via LLM.} Using the above prompt templates and few-shot examples, we employ a high-capacity LLM (e.g., GPT-4-Turbo~\cite{achiam2023gpt}) to generate candidate prompt-constraint pairs for each constraint type. By explicitly conditioning on sampled constraints and seed concepts, the LLM produces diverse, high-quality, and constraint-specific instruction tasks. This targeted generation mitigates randomness and enhances task relevance, which is essential for fine-grained evaluation of models’ instruction-following capabilities.

\parabf{Step 4: Redundancy Filtering.} 
To ensure diversity and reduce semantic redundancy, we apply post-generation filtering based on ROUGE-L similarity~\cite{lin-2004-rouge}. ROUGE-L, a widely used metric for sequence-level similarity, captures the longest common subsequence between texts, making it suitable for detecting near-duplicate prompts. Each new prompt is compared against retained ones, and those exceeding a predefined similarity threshold (0.7 in our implementations) are discarded. This filtering promotes variety, minimizes repetition, and improves coverage across constraint types. By removing similar instances, we enhance the dataset’s informativeness and robustness, better reflecting the richness of real-world constraint-driven tasks.

\parabf{Step 5: Dataset Finalization.} 
Filtered tasks undergo human review to verify correctness, clarity, and relevance. Manual validation is critical to catch subtle errors undetectable by automated scripts and to ensure the dataset's overall quality and utility for benchmarking. Only samples passing this rigorous review are included in the final dataset, guaranteeing a reliable and high-quality benchmark. 

The entire generation process can be iterated over multiple rounds until a sufficient number (e.g., $N$) of valid instructions are produced and incorporated into the benchmark.

%% file: sections/muti-level-expansion.tex
\subsection{Multi-level Instruction Expansion}
\label{sec:data_construction:multi-level}

To evaluate instruction-following under growing constraint complexity, we design a multi-level expansion framework that transforms single-level tasks into compositional sequences. Each instruction adds exactly one new constraint on top of the previous, forming a ladder from level $L1$ to $L{N}$. This allows probing the model’s ability to generalize to more complex specifications.

\parabf{Step 6: Multi-level Instruction Expansion Prompt Construction.}
Each expansion prompt evolves an existing instruction by adding one constraint. We construct a structured input where the LLM rewrites a base prompt (<base\_prompt>) into a more complex one (<prompt>) by injecting a constraint from our taxonomy. The model is guided to keep the instruction coherent, realistic, and grounded in practical software scenarios.

The prompt format largely follows initial instruction generation (Figure~\ref{fig:prompt_template}), including a constraint pool specifying the constraint type to add, a validation script, and few-shot examples. The input also includes the base prompt and target constraint type. The model must maintain realism and verifiability. An example-driven format improves output quality. Full prompt templates and examples are in our repository~\cite{replication_package}.

\parabf{Step 7: Iterative Instruction Expansion by LLM.}
From a validated base instruction, we use high-capacity LLMs (e.g., GPT-4) to iteratively generate $L2, L3, \dots, LN$ instructions via the above prompt. Each level adds one new constraint, increasing task difficulty and specificity in a controlled way. Expansion continues until level $LN$ is reached or no meaningful constraints remain. This process highlights instruction evolvability, building complex tasks progressively while preserving clarity and evaluation consistency. Figure~\ref{fig:data_format_example} shows an $L2$ task derived from an $L1$ task by adding an algorithm-type constraint.

Though constraints can theoretically combine freely, many combinations yield incoherent or non-verifiable instructions. To ensure tractability and robust evaluation, we focus on three challenging constraint types (Interface Specification, Data Structure and Code Quality from RQ2) and systematically construct multi-level variants. This enables clear performance stratification and reliable evaluation.

%% file: sections/dataset_overview.tex
\begin{figure}[htb]
	\centering
	\includegraphics[width=0.8\linewidth]{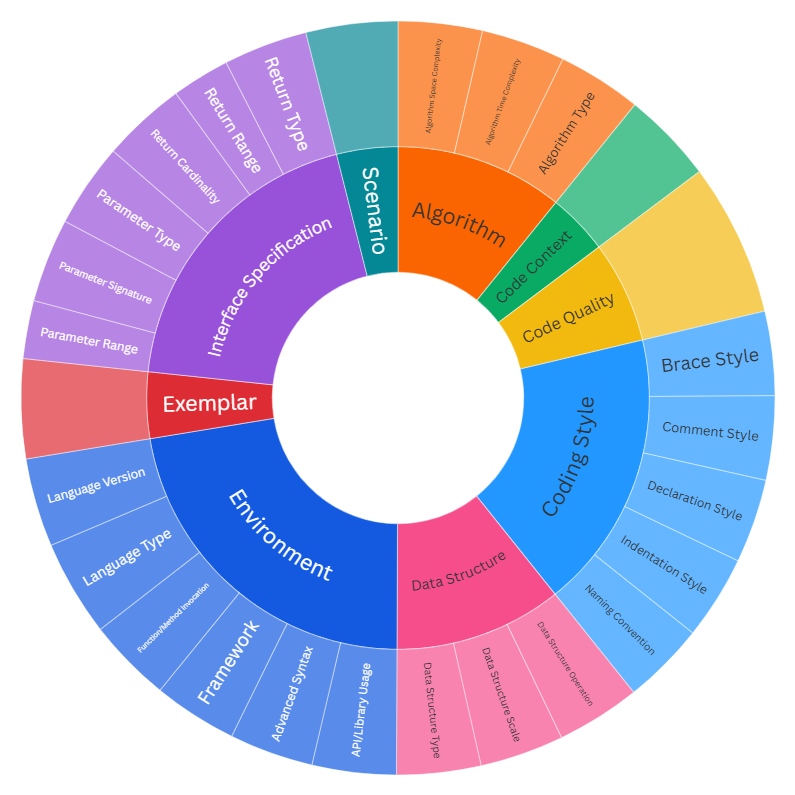}
	\caption{Overview of \bench}
	\label{fig:dataset_constraint_overview}
\end{figure}

We collected over 1,400 seed code snippets from GitHub and Hugging Face to ensure broad task coverage. Using \app, we construct \bench with GPT-4-Turbo ~\cite{achiam2023gpt} via official APIs, chosen for its strong leaderboard performance.
\bench includes 1,526 single-level and 495 multi-level constraint tasks, all explicitly conditioned on fine-grained constraints. Tasks span 9 popular languages, including Python, JavaScript, TypeScript, Java, C++, C, Go, Kotlin and C\# and 6 less popular languages, including Rust, Lua, CoffeeScript, Prolog, Julia and Haskell. Figure~\ref{fig:dataset_constraint_overview} shows the distribution of constraint counts.

Unlike prior datasets, \bench supports continuous evolution, new seeds can expand coverage, domain-specific tasks can be built via domain code, and new constraint types can be incorporated. It features more tasks, more fine-grained constraint types, and richer multi-level constraint task than existing benchmarks.

%% file: sections/experiments.tex
In this section, we comprehensively evaluate state-of-the-art LLMs on instruction-following in code generation, focusing on adherence to fine-grained constraints. Guided by our MultiCodeIF benchmark, we explore:

\begin{itemize}
    \item \textbf{RQ1 (Model Comparison)}: How do models differ in following instructions across the 9 constraint categories?

    \item \textbf{RQ2 (Constraint Impact)}: How do various constraint types affect model performance within each category?
    
    \item \textbf{RQ3 (Multi-Level Constraints)}: How well do models handle multiple simultaneous constraints with increasing difficulty?

    \item \textbf{RQ4 (Multi-Turn Evaluation)}: How effectively do models incorporate feedback to improve instruction adherence in multi-turn interactions?

\end{itemize}

\subsection{Experimental Setup}
\label{sec:experimental:setup}
\input{sections/experiments_setup}

\input{sections/rq1}

\subsection{RQ2: Constraint Impact}
\label{sec:rq2}
\input{sections/rq2}

\subsection{RQ3: Multi-Level Constraints}
\label{sec:rq3}
\input{sections/rq3}

\subsection{RQ4: Multi-Turn Evaluation}
\label{sec:rq4}
\input{sections/rq4}

%% file: sections/experiments_setup.tex
\subsubsection{Model Selection}
\label{sec:model_selection}
We selected six LLMs, including both open-source and closed-source models, covering a range of sizes and training settings. Table~\ref{tab:model_overview} presents their release dates and parameter counts. This evaluation is intended to be representative rather than exhaustive, aiming to demonstrate the effectiveness of \bench. Expanding the model set is left for future work.

\input{tables/model_overview}

Open-source models were run via the official vLLM~\cite{kwon2023efficient} framework with temperature 0.2, top-p 0.9, top-k 1, and max tokens 4096. Closed-source  models were accessed through official APIs (as of May 2025) with a max context of 8192 tokens. Prompts were standardized. Inference ran on a node with 16 vCPUs, 60 GiB RAM, and one NVIDIA A10 GPU, using Ubuntu 22.04 and Python 3.10. One output per prompt was generated to reflect typical use.


%% file: tables/model_overview.tex
\begin{table}[ht]
        \scriptsize
        \centering
        \renewcommand{\arraystretch}{0.9}
    \caption{Studied Models}

\label{tab:model_overview}
\begin{tabular}{@{}lcccc@{}}
\toprule
\textbf{Model}                      & \textbf{Release Date} & \textbf{Size} & \textbf{Context} & \textbf{Open-Source} \\ \midrule
GPT-4o \cite{hurst2024gpt}                     & Oct 2023     & -    & 128K    & $\times$    \\
Claude-3-7-Sonnet \cite{claude37sonnet2025} & Feb 2025     & -    & 200K    & $\times$    \\
DeepSeek-R1 \cite{deepseekai2025deepseekr1incentivizingreasoningcapability}          & Jan 2025     & 671B & 128K    & $\checkmark$ \\
DeepSeek-V3 \cite{deepseekai2024deepseekv3technicalreport}          & Mar 2025     & 660B & 128K    & $\checkmark$ \\
Qwen3-1.7B \cite{qwen3technicalreport}                & Apr 2025     & 1.7B & 32K     & $\checkmark$ \\
Llama-3.2-3B \cite{grattafiori2024llama}     & Sep 2024     & 3B   & 128K    & $\checkmark$ \\ \bottomrule
\end{tabular}
\end{table}

%% file: sections/rq1.tex
\subsection{RQ1: Model Comparison}
\label{sec:rq1}
We evaluate six LLMs on \bench to assess their ability to follow fine-grained, single-level constraints in code generation.

\subsubsection{Design}
\label{sec:rq1:design}
We use 1,526 single-level constraint tasks from \bench, covering all top-level categories and fine-grained constraint types. Each model is prompted with the same set of tasks under standardized conditions. After obtaining the outputs, we apply an automated evaluation pipeline to determine whether the generated code satisfies the given constraints, using strategies adapted to each constraint type (Table~\ref{tab:constraint-types}).

\begin{itemize}
    \item \textbf{Rule-based Evaluation:} For constraints verifiable via static analysis (e.g., Syntax, Data Structure), we use tools such as Tree-sitter~\cite{max_brunsfeld_2025_15531772} and Guesslang~\cite{yoeo_guesslang}.
    \item \textbf{Model-based Evaluation:} For more abstract or context-sensitive constraints (e.g., Algorithm, Code Quality), we use a strong LLM (i.e., GPT-4-Turbo) as a judge to assess alignment between the code and instruction. The evaluation prompts are publicly available in our repository~\cite{replication_package}.
\end{itemize}
The evaluation metric is accuracy, defined as the proportion of outputs that satisfy the given constraint.

\subsubsection{Results}
\label{sec:rq1:results}
\begin{figure}[t]
    \centering
    \includegraphics[width=0.9\linewidth]{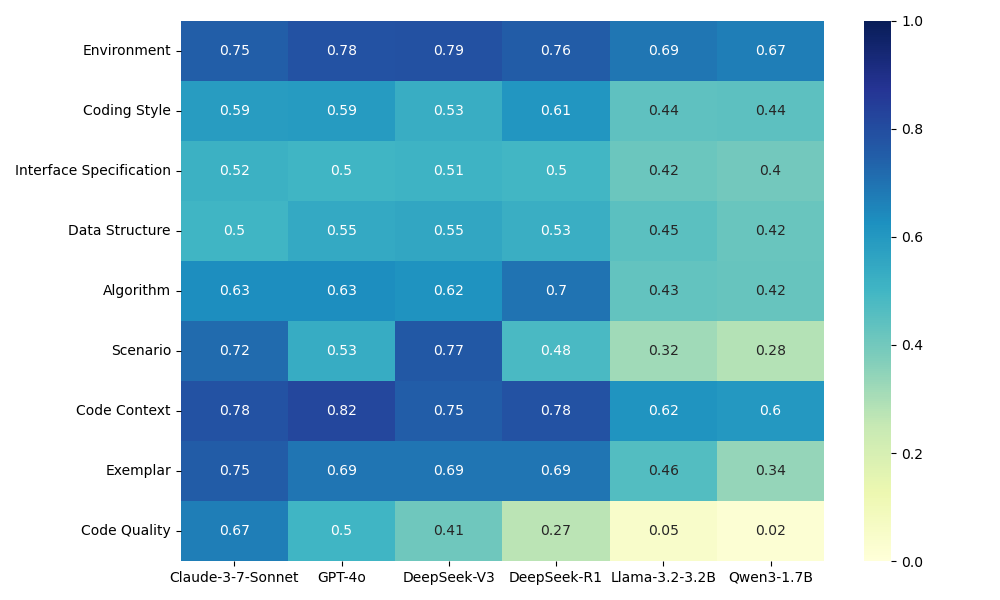}
    \caption{Constraint adherence heatmap showing model performance across 9 constraint categories. Each cell represents the adherence accuracy for a specific model-constraint pair.}
    \label{fig:constraint_model_heatmap}
\end{figure}

Table~\ref{tab:model_comp_result} and Figure~\ref{fig:constraint_model_heatmap} presents the accuracy of six LLMs across nine constraint categories. Closed-source models (Claude-3-7-Sonnet, GPT-4o) consistently outperform smaller open-source models (Llama-3.2-3B, Qwen3-1.7B) in both overall constraint adherence and specific capabilities. Claude leads with an average accuracy of 63.0\%, excelling in high-level constraints like Non-Functional Requirements (67\%) and Example (75\%). GPT-4o follows closely with 62.1\%, achieving the highest score in Code Contextual constraints (82\%). DeepSeek-V3 narrows the gap, scoring 61.3\% overall and showing strong performance in Situation (77\%) and Syntax (79\%). DeepSeek-R1 also performs well, with 60.4\% accuracy, topping Algorithm (70\%) and Coding Convention (61\%). In comparison, smaller open-source models lag behind, averaging 47\% (Llama-3.2-3B) and 45\% (Qwen3-1.7B). Their accuracy drops sharply in reasoning-intensive categories such as Non-Functional Requirements (5\% and 2\%) and Situation (32\% and 28\%), though they handle surface-level constraints like Syntax and Code Contextual reasonably well.



\input{tables/model_comp_result}

\subsubsection{Summary}
\label{sec:rq1:summary}
Instruction-following performance varies widely across models and constraint types. Large commercial models like Claude-3-7-Sonnet and GPT-4o consistently show higher adherence, especially in context-aware and non-functional constraints, reflecting better handling of complex intent. Recent open-source models like DeepSeek-V3 and R1 make notable progress, nearing commercial levels in some areas. Smaller models, however, still struggle with deep semantic understanding. These results highlight advances in strong models but also persistent gaps for resource-limited systems.

%% file: tables/model_comp_result.tex
\begin{table*}[]
\footnotesize
    \centering
    \caption{Accuracy of Six LLMs on Single-Level Instruction-Following Tasks from \bench Across 9 Constraint Categories}
\label{tab:model_comp_result}
\renewcommand{\arraystretch}{1}
\begin{tabular}{l|cccccccccc}
\hline
\textbf{Model} &
  \textbf{Environment} &
  \textbf{\begin{tabular}[c]{@{}c@{}}Coding\\ Style\end{tabular}} &
  \textbf{\begin{tabular}[c]{@{}c@{}}Interfac\\ Specification\end{tabular}} &
  \textbf{\begin{tabular}[c]{@{}c@{}}Data\\ Structure\end{tabular}} &
  \textbf{Algorithm} &
  \textbf{\begin{tabular}[c]{@{}c@{}}Code\\ Quality\end{tabular}} &
  \textbf{Scenario} &
  \textbf{\begin{tabular}[c]{@{}c@{}}Code\\ Context\end{tabular}} &
  \textbf{Exemplar} &
  \textbf{Avg.} \\ \hline
Claude-3-7-Sonnet &
  74.8 &
  58.8 &
  \textbf{51.7} &
  50.3 &
  \textbf{63.0} &
  \textbf{67.0} &
  71.7 &
  78.3 &
  \textbf{75.4} &
  \textbf{63.0} \\
GPT-4o &
  78.3 &
  59.1 &
  50.0 &
  54.5 &
  63.0 &
  50.0 &
  59.1 &
  \textbf{81.7} &
  69.2 &
  62.1 \\
DeepSeek-V3 &
  \textbf{78.9} &
  52.9 &
  51.0 &
  \textbf{55.2} &
  61.8 &
  41.0 &
  \textbf{76.7} &
  75.0 &
  69.2 &
  61.3 \\
DeepSeek-R1 &
  75.7 &
  \textbf{60.6} &
  49.7 &
  52.7 &
  \textbf{69.7} &
  27.0 &
  48.3 &
  78.3 &
  69.2 &
  60.4 \\
Llama-3.2-3B &
  68.9 &
  43.8 &
  41.6 &
  44.8 &
  43.0 &
  5.0 &
  31.7 &
  61.7 &
  46.2 &
  46.8 \\
Qwen3-1.7B &
  66.9 &
  44.2 &
  39.9 &
  41.8 &
  42.4 &
  2.0 &
  28.3 &
  60.0 &
  33.8 &
  44.8 \\ \hline
\end{tabular}
\end{table*}

%% file: sections/rq2.tex
To understand how constraint types affect instruction-following in code generation, we conduct a constraint-focused analysis in RQ2. This reveals which instructions are more challenging and where LLMs commonly fail.

\subsubsection{Design}
\label{sec:rq2:design}
In RQ2, we shift from comparing models to comparing constraint categories. Our aim is to see how constraint nature impacts LLMs’ instruction adherence. We identify which constraints are easier or harder across models by averaging accuracy over all six models from RQ1. This model-agnostic approach isolates the impact of constraint semantics on adherence rates.


\subsubsection{Results}
\label{sec:rq2:results}

\begin{figure}[t]
    \centering
    \includegraphics[width=0.9\linewidth]{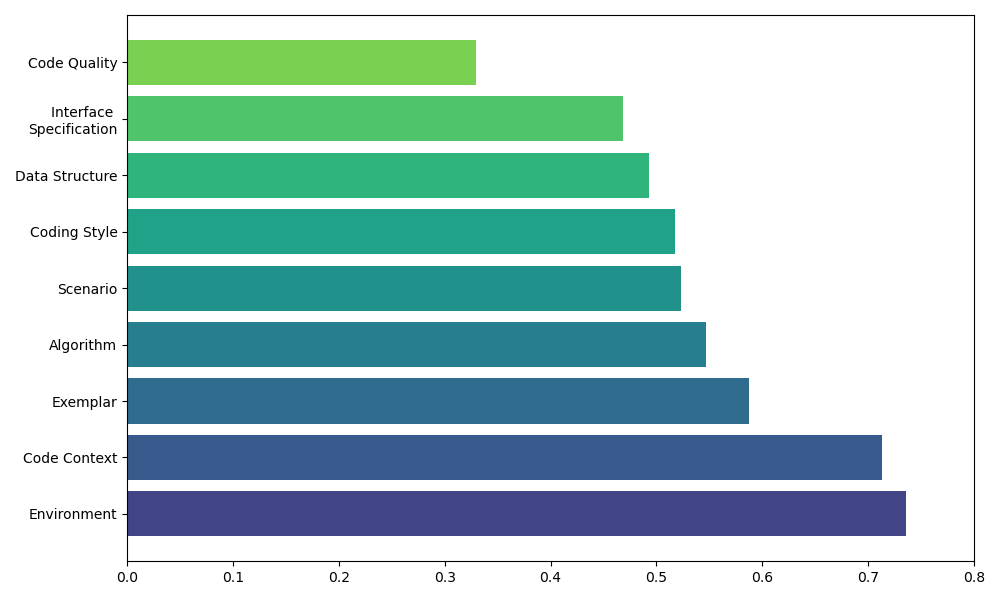}
    \caption{Average Accuracy for Each Constraint Category across All Six Models}
    \label{fig:model_avg}
\end{figure}

Figure~\ref{fig:model_avg} shows the average accuracy for each constraint category across all six models. A clear hierarchy of difficulty emerges among the 9 categories.

\begin{itemize}
    \item \textbf{High Accuracy:} \textbf{Environment} (73.6\%), \textbf{Code Context} (71.3\%), and \textbf{Exemplar} (58.8\%) rank highest. Environment constraints are easily verifiable via parsing, while Exemplar constraints benefit from models’ in-context learning and pattern continuation skills.

    \item \textbf{Moderate Accuracy:} \textbf{Algorithm} (54.7\%), \textbf{Scenario} (52.3\%), and \textbf{Coding Style} (51.8\%) require both surface-level recognition and deeper understanding. Success often depends on whether the task aligns with training data patterns.

    \item \textbf{Low Accuracy:} \textbf{Interface Specification} (46.8\%), \textbf{Data Structure} (49.3\%), and especially \textbf{Code Quality} (33.0\%) demand abstract reasoning or attention to implicit aspects. Non-functional constraints like performance or readability are frequently overlooked.


\end{itemize}

\begin{figure}[t]
    \centering
    \includegraphics[width=0.7\linewidth]{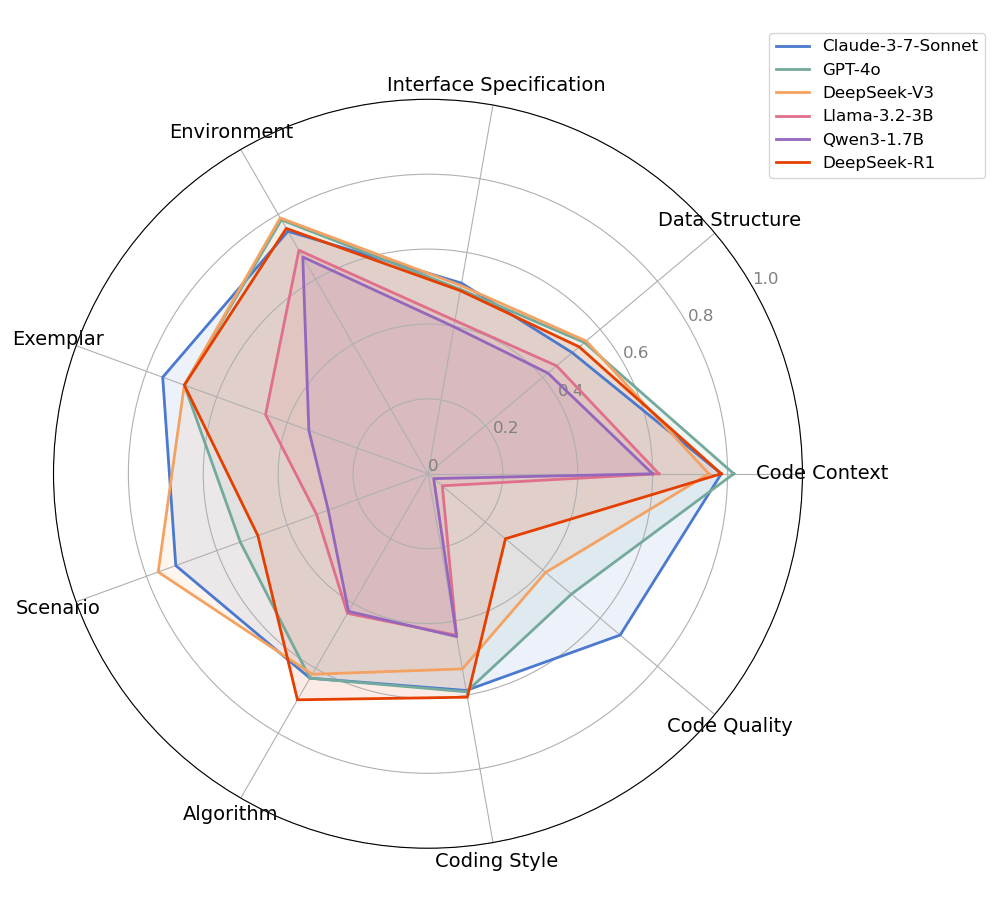}
    \caption{Constraint adherence distribution across 9 categories for each model.}
    \label{fig:radar_constraint_models}
\end{figure}

As shown in Figure~\ref{fig:radar_constraint_models}, constraint adherence also varies notably between closed-source large models and open-source small models. Closed-source models outperform across all categories, with the largest gap in \textbf{Code Quality}, achieving 52.7\% accuracy versus just 3.5\% for open-source models. This highlights a broader pattern: abstract, non-functional constraints remain poorly handled, particularly by smaller models.

\subsubsection{Summary}
\label{sec:rq2:summary}
Our constraint-centric analysis reveals clear differences in instruction-following difficulty across constraint types, regardless of model architecture. Explicit, surface-level constraints like \textbf{Environment}, \textbf{Code Context}, and \textbf{Exemplar} show high adherence, reflecting LLMs’ strengths in pattern matching and local context use. In contrast, \textbf{Code Quality}, \textbf{Interface Specification}, and \textbf{Data Structure} show low adherence, exposing challenges with abstract, implicit, or global code properties. This perspective offers a diagnostic tool for model evaluation and points to the need for better instruction encoding and semantic reasoning.

%% file: sections/rq3.tex
To further probe LLMs' capability in handling compositional instructions, we introduce a multi-level constraint composition setting. Unlike RQ1 and RQ2, which evaluate models on isolated, single-level constraints, this experiment examines how well models perform when multiple constraints are combined within a single instruction. Our key goal is to assess whether LLMs can maintain or improve adherence as constraints become more numerous, specific, and demanding.

\subsubsection{Design}
We evaluate models using the 495 multi-level tasks from \bench, where each task is structured into three levels of increasing constraint complexity ($L2$ to $L4$). While $L1$ contains a single, base-level constraint, levels $L2$ to $L4$ progressively add more constraints, forming increasingly compositional instructions. To assess how constraint composition affects instruction adherence, we compare model performance across all levels - $L1$ to $L4$ - thereby capturing how adherence degrades or holds up as constraints accumulate. Previous experiments showed DeepSeek-R1 slightly underperforms DeepSeek-V3 in instruction adherence and has slower inference speed; for efficiency, we conduct experiments only on the other five models.

Following FollowBench \cite{jiang2023followbench}, we adopt two complementary metrics to evaluate instruction adherence under constraint composition:

\begin{itemize}
    \item \textbf{SSR (Soft Satisfaction Rate).} SSR reflects the proportion of all individual constraints satisfied by the model's output, treating each constraint equally regardless of its level. It provides a fine-grained, level-agnostic view of instruction adherence. Formally:
    \[
    \textstyle
    \text{SSR} = \frac{1}{|\mathcal{C}|} \sum_{c \in \mathcal{C}} \mathbb{I}[\text{sat}(c)]
    \]
    where $\mathcal{C}$ is the set of all constraints in a composed instruction, and $\mathbb{I}[\text{sat}(c)]$ is an indicator function that returns 1 if constraint $c$ is satisfied, and 0 otherwise.

    \item \textbf{HSR (Hard Satisfaction Rate).} HSR evaluates whether a model satisfies \textit{all} constraints within a given instruction. It returns 1 only if the output satisfies every constraint; otherwise, it returns 0. This metric captures the stricter, all-or-nothing interpretation of instruction-following. Formally:
    \[
    \textstyle
    \text{HSR} = \mathbb{I}\left[\bigwedge_{c \in \mathcal{C}} \text{sat}(c)\right]
    \]
\end{itemize}

\subsubsection{Results}
\begin{figure}[t]
    \centering
    \includegraphics[width=0.6\linewidth]{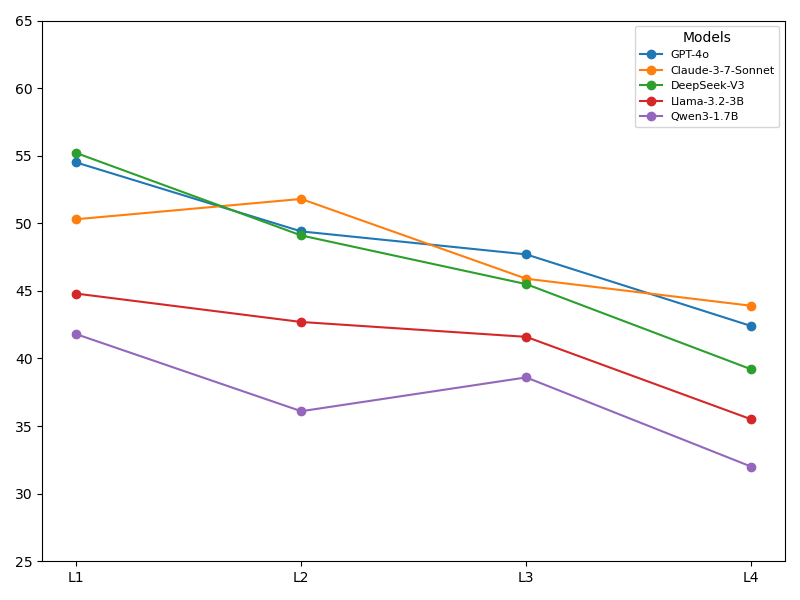}
    \caption{Model Performance Across Constraint Complexity Levels.}
    \label{fig:multi_level_result}
\end{figure}

Table~\ref{tab:multi_level_result} and Figure~\ref{fig:multi_level_result} report the multi-level constraint SSR and HSR across five LLMs, specifically within the \textbf{Data Structure} constraint category. In this setting, each level from \textbf{$L1$} to \textbf{$L4$} progressively introduces new structural constraints, for example, requiring the use of specific data structure types, enforcing data processing methods, and mandating elements size. Several trends emerge:

\begin{itemize}
\item \textbf{Decreasing Adherence with Constraint Depth.} All models show a marked decline in adherence as more data structure constraints are imposed. This confirms that reasoning over increasingly complex structural requirements, such as managing multiple nested data types or adhering to strict access semantics, poses a significant challenge for current LLMs. On average, performance drops by 10-15 percentage points from $L1$ to $L4$.

\item \textbf{Larger Models are More Stable.} Models such as GPT-4o (48.5\%), Claude-3-7-Sonnet (48.0\%), and DeepSeek-V3 (47.3\%) maintain relatively high average adherence scores, even at $L4$. These models demonstrate better structural generalization and robustness when composing multiple constraints. In contrast, smaller models like Llama-3.2-3B and Qwen3-1.7B fall below 42\% and 38\%, respectively.

\item \textbf{Local Improvements from Constraint Clarification.} Claude-3-7-Sonnet shows a performance increase from $L1$ to $L2$, suggesting that in certain cases, adding structural constraints can disambiguate vague base tasks. This indicates that constraints do not universally increase difficulty, some may provide guiding inductive bias that benefits generation.

\item \textbf{Compression of Model Differences at Higher Levels.} By $L4$, model performance converges more closely, narrowing the gap between stronger and weaker models. This suggests that multi-level structural constraint composition acts as a leveling force, revealing fundamental limitations in architectural generalization capabilities across the board.
\end{itemize}

\input{tables/multi_level_result}

\subsubsection{Summary}
\label{sec:rq3:summary}
Multi-level constraints surface the limits of compositional reasoning in LLMs. This reflects a core limitation in compositional instruction-following: models often succeed on isolated constraints but fail to integrate them reliably. Our multi-level setup provides a scalable path to probe and extend this behavior.

%% file: tables/multi_level_result.tex
\begin{table*}[]
\centering
\renewcommand{\arraystretch}{0.9}
\scriptsize
\caption{Multi-level adherence scores for Data Structure constraints}
\label{tab:multi_level_result}
\begin{tabular}{@{}l|ccccc|ccccc@{}}
\toprule
\multirow{2}{*}{\textbf{Model}} &
  \multicolumn{5}{c|}{\textbf{SSR (\%)}} &
  \multicolumn{5}{c}{\textbf{HSR (\%)}} \\ \cmidrule(l){2-11} 
 &
  \textbf{L1} &
  \textbf{L2} &
  \textbf{L3} &
  \textbf{L4} &
  \textbf{Avg.} &
  \textbf{L1} &
  \textbf{L2} &
  \textbf{L3} &
  \textbf{L4} &
  \textbf{Avg.} \\ \midrule
GPT-4o            & 54.5 & 49.4 & 47.7 & 42.4 & 48.5 & 54.5 & 35.2 & 26.1 & 18.8 & 33.7 \\
Claude-3-7-Sonnet & 50.3 & 51.8 & 45.9 & 43.9 & 48.0 & 50.3 & 37.6 & 24.8 & 20.6 & 33.3 \\
DeepSeek-V3       & 55.2 & 49.1 & 45.5 & 39.2 & 47.3 & 55.2 & 35.8 & 23.6 & 14.5 & 32.3 \\
Llama-3.2-3B      & 44.8 & 42.7 & 41.6 & 35.5 & 41.2 & 44.8 & 30.9 & 20.6 & 13.9 & 27.5 \\
Qwen3-1.7B        & 41.8 & 36.1 & 38.6 & 32.0 & 37.1 & 41.8 & 21.8 & 16.4 & 10.9 & 22.7 \\ \bottomrule
\end{tabular}
\end{table*}

%% file: sections/rq4.tex
In this RQ, we examine whether LLMs can improve instruction adherence via multi-turn interactions. Specifically, a model generates an initial solution, receives feedback on violated constraints, and revises its output accordingly (i.e., a self-repair process). We assess how well models incorporate feedback across constraint types and complexity levels to evaluate their iterative improvement.

\begin{figure}
    \centering
    \includegraphics[width=\linewidth]{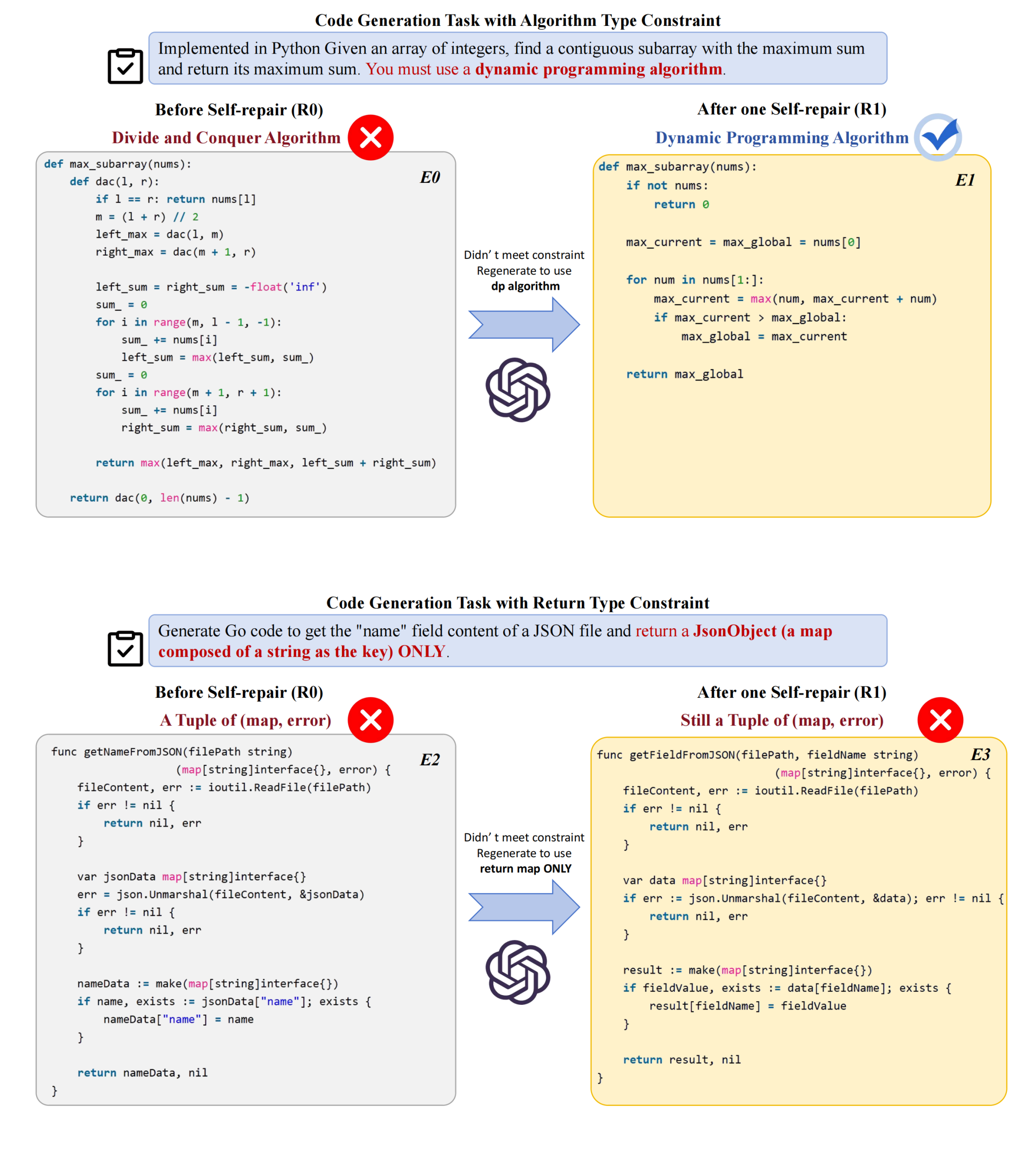}
    \caption{Example of self-repair that could lead to two different results. One is that the constraints are met after repair, and the other is that the constraints are still not met after repair.}
    \label{fig:self_repair_example}
\end{figure}

\subsubsection{Design} The experiment involved testing selected model on \app using controlled prompts, with output correctness assessed through automatic evaluation. The experimental design is structured as follows:

\textbf{Evaluation Process.} To assess the potential of self-repair strategies, we implement a looped mechanism where a model iteratively revises its own outputs in response to constraint violation feedback. At iteration $k$, if the model output fails to satisfy all constraints, we construct a follow-up input for iteration $k{+}1$ containing the original instruction along with structured diagnostics highlighting the specific unmet constraints. The model then attempts to produce a corrected version. Each instance undergoes up to $N$ rounds of repair. We focus on two high-performing models, Claude-3-7-Sonnet and GPT-4o, selected based on their strong results in RQ1.

\begin{figure}
    \centering
    \includegraphics[width=0.65\linewidth]{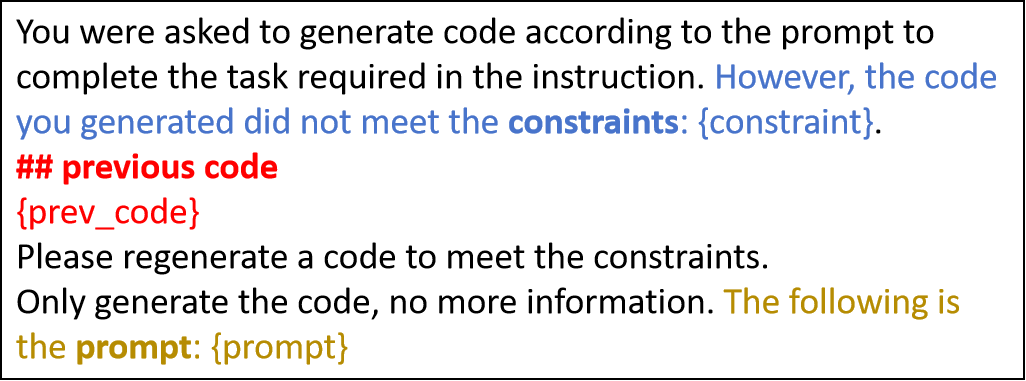}
    \caption{Self-repair prompt template used to automatically regenerate code after constraint violations.}
    \label{fig:self_repair_prompt}
\end{figure}

\textbf{Metrics: IFRepair@\textit{k}.} To quantify the effectiveness of iterative self-repair under constraint feedback, we define \textbf{IFRepair@\textit{k}} as the average Hard Satisfaction Rate (HSR) after the $k$-th round of repair. Formally, let $\mathcal{D} = \{x_1, x_2, \ldots, x_n\}$ denote the dataset of $n$ instruction instances, where each instance $x_i$ is associated with a constraint set $C_i$. Let $y_i^{(k)}$ be the model output after the $k$-th self-repair iteration for input $x_i$. Then:

\[
\textstyle
\text{IFRepair@}k = \frac{1}{n} \sum_{i=1}^{n} \mathbb{I}\left[\bigwedge_{c \in C_i} \text{sat}(c, y_i^{(k)})\right]
\]




This formulation captures the model’s instruction-following success rate at iteration $k$, reflecting how well the model leverages diagnostic feedback to revise its outputs. A higher IFRepair@\textit{k} value indicates greater compliance with fine-grained constraints after $k$ rounds of self-repair. To assess improvement, we also define the per-round gain:

\[
\textstyle
\Delta_k = \text{IFRepair@}k - \text{IFRepair@}(k-1)
\]

This delta reflects the effectiveness of each feedback iteration in promoting complete constraint adherence.

\subsubsection{Results}
Table~\ref{tab:self_repair} presents the IFRepair@\textit{k} scores across five iterations ($k{=}0$ to $4$) for both Claude-3-7-Sonnet and GPT-4o. The baseline performance without repair ($k{=}0$) is 63.0\% for Claude and 62.1\% for GPT-4o. One round of self-repair yields significant gains, 71.2\% for Claude and 70.6\% for GPT-4o-corresponding to improvements of 8.2 and 8.5 percentage points, respectively. This highlights that structured feedback is immediately beneficial for both models.
Subsequent iterations show continued, though diminishing, improvements. Claude’s performance rises to 76.5\% at $k{=}2$, 82.3\% at $k{=}3$, and 83.4\% at $k{=}4$, with gains of +5.3\%, +5.8\%, and +1.1\% per round. GPT-4o exhibits a similar trend: 74.2\% at $k{=}2$, 77.6\% at $k{=}3$, and 78.6\% at $k{=}4$, with gains of +3.6\%, +3.2\%, and +1.0\%. While GPT-4o starts slightly behind Claude, the gap narrows by the final iteration.

\input{tables/self_repair_result}

\subsubsection{Summary}
The results of RQ4 confirm that structured self-repair substantially improves instruction-following under fine-grained constraints. Over 4 repair rounds, Claude-3-7-Sonnet and GPT-4o improve from 63.0\% to 83.4\% and from 62.1\% to 78.6\%, respectively. Most gains occur in the first two rounds, indicating that LLMs can quickly incorporate explicit feedback. However, diminishing returns after the third iteration suggest limits due to model capacity or task difficulty. Overall, self-repair proves to be an effective strategy for addressing instruction-following gaps. These findings underscore the value of feedback-aware workflows and point to the potential of iterative refinement in future model training.

\begin{figure}
	\centering
	\includegraphics[width=\linewidth]{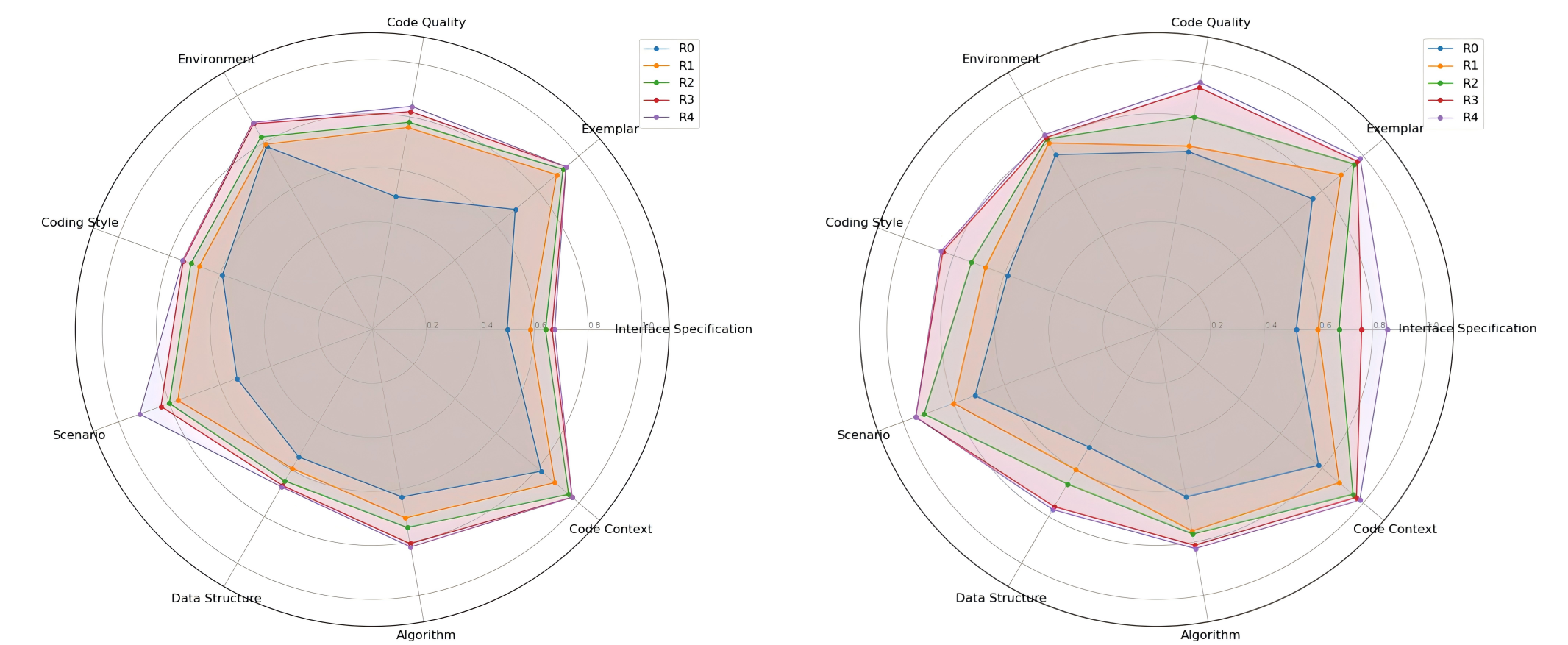}
	\caption{Performance of Claude 3-7-Sonnet and GPT-4o across constraint categories over four self-repair iterations (R0-R4).}
	\label{fig:self_repair_radar}
\end{figure}

%% file: tables/self_repair_result.tex
\begin{table}[h]
\centering
\scriptsize
\renewcommand{\arraystretch}{0.75}
\caption{IFRepair@\textit{k} across self-repair iterations using Claude-3-7-Sonnet and GPT-4o.}
\label{tab:self_repair}
\begin{tabular}{@{}c|cc|cc@{}}
\toprule
\multirow{2}{*}{\textbf{Round ($k$)}} & \multicolumn{2}{c|}{\textbf{Claude-3-7-Sonnet}} & \multicolumn{2}{c}{\textbf{GPT-4o}}     \\ \cmidrule(l){2-5} 
                                      & \textbf{IFRepair@\textit{k}}     & \textbf{Delta}     & \textbf{IFRepair@\textit{k}} & \textbf{Delta} \\ \midrule
$k=0$ (Initial) & 63.0\% & --   & 62.1 & --   \\
$k=1$           & 71.2\% & +8.2\% & 70.6\% & +8.5\% \\
$k=2$          & 76.5\% & +5.3\% & 74.2\% & +3.6\% \\
$k=3$           & 82.3\% & +5.8\% & 77.6\% & +3.2\% \\
$k=4$           & 83.4\% & +1.1\% & 78.6\% & +1.0\% \\ \bottomrule
\end{tabular}
\end{table}

%% file: sections/related_work.tex
Many benchmarks have been introduced to evaluate LLMs’ code generation abilities: HumanEval~\cite{chen2021evaluatinglargelanguagemodels}, MBPP~\cite{austin2021programsynthesislargelanguage}, 
and DS-1000~\cite{DS-1000} target function-level tasks; ClassEval~\cite{classeval} focuses on class-level generation; CoderEval~\cite{codereval} and DevEval~\cite{deveval} assess performance at the repository level for practical engineering scenarios. In addition, RustEvo\textsuperscript{2}~\cite{liang2025rustevo} introduces a benchmark targeting API evolution scenarios in Rust, while FeedbackEval~\cite{dai2025feedbackeval} evaluates LLMs under feedback-driven code repair settings.

However, the instruction-following ability of LLMs in the domain of code generation remains underexplored. In the field of natural language processing , several benchmarks, such as IFEval~\cite{IFEval}, InFoBench~\cite{InFoBench}, and ComplexBench~\cite{ComplexBench}, have already been developed to assess instruction-following capabilities. Recently, a few benchmarks have emerged to evaluate LLMs' instruction-following performance in code generation, such as CodeIF~\cite{CodeIF} and CodeIF-Bench~\cite{CodeIF-Bench}. Nevertheless, these benchmarks suffer from several limitations, including coarse-grained constraints, limited language support, static task definitions, and a lack of instructions related to iterative refinement or self-correction. Moreover, these benchmarks often overlook scenarios involving semantic preservation and multi-level instruction alignment. Recent work on equivalent code representations~\cite{li2024generating} demonstrates the potential of LLMs to generate constrained or unconstrained semantic alternatives (e.g., pseudocode, comments), which can complement instruction-based generation. Similarly, CodeScore~\cite{10.1145/3695991} proposes a functionality-centered evaluation metric that overcomes the surface-level limitations of traditional match-based metrics, offering finer-grained assessment of instruction adherence and execution correctness. To address these limitations, we propose \bench, a benchmark designed to assess the instruction-following capabilities of LLMs specifically in the context of code generation.

Since acquiring real-world instruction data is often expensive and labor-intensive, automatic instruction generation has gained attention as a feasible alternative. Self-Instruct~\cite{Self-Instruct} employs PLMs to generate instruction data from a small seed set. Evol-Instruct~\cite{Evol-Instruct} introduces mutations to meta-instructions to create more complex and diverse tasks, thereby increasing reasoning depth and constraint complexity. OSS-Instruct~\cite{OSS-Instruct} and WaveCoder~\cite{WaveCoder} utilize LLMs to generate code-related instruction data, while Genetic-Instruct~\cite{Genetic-Instruct} combines classical genetic algorithms with LLMs to generate more diverse code task instructions. In contrast to previous code instruction generation approaches, \app automatically synthesizes multi-level constraints based on predefined constraint types, and subsequently leverages LLMs to generate diverse code task instructions guided by these constraints. All generated instructions are manually verified to ensure their correctness for evaluations.

%% file: sections/threats.tex
Although our dataset is automatically generated using LLMs, which may introduce noise, we mitigate this risk by seeding tasks from real-world code, applying similarity-based filtering to remove duplicates, and performing manual validation to ensure correctness and diversity. While our model selection is limited, we include six representative state-of-the-art models, such as DeepSeek-R1, to demonstrate that \bench{} provides a reliable basis for evaluating instruction adherence. To ensure consistency, we use only official model releases. To address concerns about generalizability, \bench{} spans fourteen programming languages (e.g., Python, Java, C++, Rust), broadening its applicability. Finally, our benchmark includes over 1,500 single-constraint and nearly 500 multi-constraint tasks, exceeding the scale of prior datasets. Together, these efforts help minimize validity threats and establish \bench{} as a robust and extensible evaluation framework for instruction-following in code generation.

%% file: sections/conclusion.tex
In this work, we introduced \bench{}, a hierarchical, multi-language benchmark for evaluating instruction adherence of large language models in code generation tasks with fine-grained, compositional constraints. At its core is a constraint taxonomy with 9 categories and 27 fine-grained types, supporting systematic evaluation across functional and non-functional dimensions. To build \bench{} at scale, we developed \app{}, an automated pipeline that uses real-world code, prompt-based generation, similarity filtering, and manual validation to synthesize and evolve tasks. Experiments on six leading models reveal significant performance gaps across constraint types and model scales, with abstract and non-functional constraints posing persistent challenges. Multi-turn evaluation shows that feedback-guided refinement can substantially boost adherence. With its scalable task construction, diverse constraint coverage, and interactive evaluation, \bench{} provides a solid foundation for advancing reliable and controllable code generation. Future work will expand model coverage and explore richer feedback strategies for improved generalization.